\documentclass[prl,twocolumn,showpacs,amsmath,amssymb,superscriptaddress]{revtex4}
\usepackage{graphicx} 

\newcommand{\rmd}{\mathrm{d}}
\newcommand{\rmF}{\mathrm{F}}
\newcommand{\bfr}{\mathbf{r}}

\newcommand{\bfF}{\mathbf{F}}

\begin{document}

\title{Laser-induced bound-state phases in high-order harmonic generation}

\author{Adam Etches}
\affiliation{Lundbeck Foundation Theoretical Center for Quantum System Research, Department of Physics and Astronomy, Aarhus University, 8000 Aarhus C, Denmark}

\author{Mette B. Gaarde}
\affiliation{Department of Physics and Astronomy, Louisiana State University, Baton Rouge, Louisiana 70803-4001, USA}

\author{Lars Bojer Madsen}
\affiliation{Lundbeck Foundation Theoretical Center for Quantum System Research, Department of Physics and Astronomy, Aarhus University, 8000 Aarhus C, Denmark}

\date{\today}

\begin{abstract}
We present single-molecule and macroscopic calculations showing that laser-induced Stark shifts contribute significantly to the phase of high-order harmonics from polar molecules. This is important for orbital tomography, where phases of field-free dipole matrix elements are needed in order to reconstruct molecular orbitals. We derive an analytical expression that allows the first-order Stark phase to be subtracted from experimental measurements.
\end{abstract}


\maketitle

It is a long-standing goal of atomic and molecular physics to follow electronic processes on their natural length and time scales, and eventually even to control their dynamics. Advances towards orbital tomography, the experimental reconstruction of an electronic wave function, have been made using high-order harmonic generation (HHG)~\cite{itatani:867, haessler:200, vozzi:822}. In HHG based tomography, a continuum electron wave packet is formed at the peak of each half-cycle of an infrared driving laser. The continuum wave packet is accelerated in the oscillating laser field, and brought back to the vicinity of the bound-state wave packet at a later time. A high-energy photon is emitted when the continuum electron recombines into the ground state after having picked up kinetic energy in the laser field~\cite{krause:3535, corkum:1994}. The recombination step encodes information about the ground state onto the emitted harmonics through the recombination dipole matrix elements. 

In order to reconstruct the ground state orbital, one needs both the magnitudes and the phases of the recombination matrix elements. It is challenging to extract these phases from HHG spectra, as phases not related to the recombination step have to be subtracted from the harmonic phases. Phases related to the formation of a continuum wave packet and its subsequent acceleration in the laser field have already been accounted for, allowing the reconstruction of symmetric orbitals in N$_2$ and CO$_2$~\cite{haessler:200, vozzi:822}. Nonsymmetric orbitals, however, pose a problem for orbital tomography due to their permanent dipoles, which cause them to acquire an additional laser-induced bound-state Stark phase. Using CO as a representative polar molecule, we find that the first-order Stark phase grows from zero to $0.5 \pi$ within the harmonic plateau. The aim of this work is to extend the tomographic method by accounting for this Stark phase, and presenting an analytical expression with which to subtract it from measurements. 

\begin{figure}
 \begin{center}
   \includegraphics[width=\columnwidth]{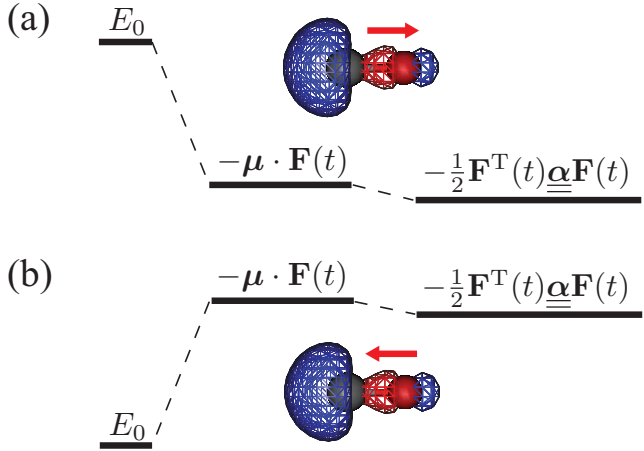}
 \end{center}
 \caption{(Color online) Sketch of the Stark shift felt by the HOMO of CO when subject to an electric field. The relative size of the first- and second-order Stark shift is drawn to scale for an electric field $\rmF_0 = 0.075$~au, corresponding to a laser intensity of $2 \times 10^{14}$~W$/$cm$^2$. (a)~The electric field is parallel to the permanent dipole of the orbital, and the effective ionization potential is raised. (b)~The electric field is antiparallel to the permanent dipole, and the effective ionization potential is lowered.}
 \label{Fig1}
\end{figure}

To isolate the effect of Stark phases we consider the idealized~\cite{stapelfeldt:543, holmegaard:023001, filsinger:064309, ghafur:289, de:153002} case of perfectly 3D oriented molecules. In the simplest model, the electric field $\bfF(t)$ of the driving pulse only influences the highest occupied molecular orbital (HOMO) by adiabatically Stark-shifting its orbital energy $E$ (atomic units are used throughout):
\begin{align}
  \label{E(F)}
  E(\bfF(t)) = E_0 - \boldsymbol{\mu} \cdot \bfF(t) - \frac{1}{2} \bfF^{\mathrm{T}}(t) \underline{\underline{\boldsymbol{\alpha}}} \bfF(t).
\end{align}
As an example, we take CO oriented parallel to the polarization axis of the driving laser. The HOMO has $\sigma$ symmetry, and an ionization potential of $I_{\mathrm{p}} = |E_0| = 0.5150$~au. Its permanent dipole $|\boldsymbol{\mu}| = 1.1$~au points from the carbon towards the oxygen nucleus. The components of the polarizability tensor $\underline{\underline{\boldsymbol{\alpha}}}$ are $\alpha_{\parallel} = 3.2$~au, and $\alpha_{\perp} = 2.8$~au~\cite{etches:155602}. The resulting Stark shift caused by a moderately intense laser is sketched in Fig.~\ref{Fig1}. The chosen geometry suppresses contributions from the HOMO$-1$ due to its $\pi$ symmetry~\cite{madsen:043419}.

As the bound state evolves in time from $t'$ to $t$, it accumulates the phase $\int_{t'}^t E(\bfF(t'')) \rmd t''$. According to Eq.~\eqref{E(F)} the accumulated phase differs from the field-free time-evolution by a Stark phase 
\begin{align}
  \label{Phi tot}
  \Phi_{\mathrm{Stark}}(t,t') = \Phi_{\mathrm{Stark}}^{(1)}(t,t') + \Phi_{\mathrm{Stark}}^{(2)}(t,t'),
\end{align}
where the first- and second-order Stark phases read
\begin{align}
  \label{Phi1}
  \Phi_{\mathrm{Stark}}^{(1)}(t,t') &= -\int_{t'}^t \boldsymbol{\mu} \cdot \bfF(t'') \rmd t'' \\
  \Phi_{\mathrm{Stark}}^{(2)}(t,t') &=  - \frac{1}{2} \int_{t'}^t \bfF^{\mathrm{T}}(t'') \underline{\underline{\boldsymbol{\alpha}}} \bfF(t'') \rmd t''.
  \label{Phi2}
\end{align}
The Stark phase changes the interference between the bound state, and the continuum wave packet that is born at time $t'$, and returns at time $t$. 

\begin{figure}
 \begin{center}
   \includegraphics[width=\columnwidth]{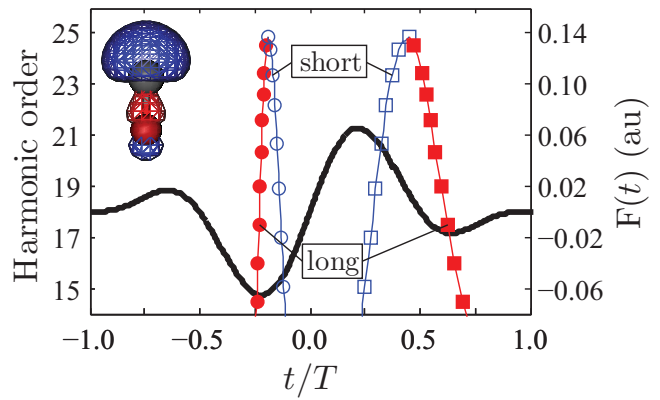}
 \end{center}
 \caption{(Color online) Classical trajectories for CO subject to an ultrashort laser pulse, indicated by the solid black curve. Instants of ionization are marked by circles, and instants of recombination by squares. The vertical offset of a trajectory, consisting of a circle and a square, indicates the photon energy of the emitted harmonic, calculated using Eq.~\eqref{omegaHHG}. Open (blue) markers indicate short trajectories, and full (red) markers indicate long trajectories.}
  \label{Fig2}
\end{figure}

Equations~\eqref{Phi tot}--\eqref{Phi2} are formulated in the time-domain, while orbital tomography is performed in the frequency domain. We therefore calculate classical electron trajectories, and translate the kinetic energy into photon energy using energy conservation for those continuum electrons that recombine at the origin:
\begin{align}
  \label{omegaHHG}
  \omega(t,t') = \frac{1}{2} \dot{\mathrm{r}}^2(t,t') + I_{\mathrm{p}}(\bfF(t)).
\end{align}
Here $\dot{\bfr}(t,t')$ is the velocity at time $t$ of an electron that is detached at the origin with zero velocity at time $t'$, and $I_{\mathrm{p}}(\bfF(t)) = |E(\bfF(t))|$. The effect of the ionic potential on the continuum trajectories is ignored. The result of a classical trajectory calculation is shown in Fig.~\ref{Fig2} for an ultrashort pulse. The electric field is assumed to have a cosine squared envelope, and an 800~nm sine carrier wave. The full width at half maximum is $0.72$ times one optical cycle $T$, corresponding to a full duration of $2T$. The maximum of the electric field is $\rmF_0 = 0.071$~au, corresponding to a peak intensity of $2.0 \times 10^{14}$~W$/$cm$^2$ of the envelope. Only one half-cycle contributes to the harmonic emission, which ensures that the continuum electron only recollides from one direction. Controlling the recollision direction is essential for tomography of nonsymmetric states~\cite{zwan:033410}. Together, Eq.~\eqref{omegaHHG} and the calculation leading to Fig.~\ref{Fig2} provide the required map between $(t,t')$ and harmonic frequency. 

Returning to the Stark phases, Eq.~\eqref{Phi1} can be interpreted in terms of the integral of the force felt by the continuum electron, showing that the first-order Stark phase is directly proportional to the return velocity
\begin{align}
  \label{Phi1 simplified}
  \Phi_{\mathrm{Stark}}^{(1)}(t,t') &= \boldsymbol{\mu} \cdot \dot{\bfr}(t,t').
\end{align}
With orbital tomography in mind, one might be interested in minimizing the Stark phase. However, Eq.~\eqref{Phi1 simplified} reveals that the first-order Stark phase cannot be minimized by varying the driving pulse, as it only depends on the return velocity, and the angle $\theta$ between the internuclear axis and the laser polarization axis. Insertion into Eq.~\eqref{omegaHHG} gives
\begin{align}
  \label{Phi1 time-dependent model}
  \Phi_{\mathrm{Stark}}^{(1)}(t,t') = & \mathrm{sgn} \left( \boldsymbol{\mu} \cdot \dot{\bfr}(t,t') \right) \mu \left| \cos(\theta) \right| \nonumber \\
  & \times \sqrt{2\left( \omega(t,t') - I_{\mathrm{p}}(\bfF(t)) \right)},
\end{align}
where $\mathrm{sgn}(\boldsymbol{\mu} \cdot \dot{\bfr}(t,t'))$ keeps track of the direction with which the returning electron probes the bound state. The time-dependence of the Stark-shifted ionization potential introduces a small difference between short and long trajectories. If the field-dependence of the ionization potential is ignored, then Eq.~\eqref{Phi1 time-dependent model} simplifies to
\begin{align}
  \label{Phi1 model}
  \Phi_{\mathrm{Stark}}^{(1)}(\omega) \approx \pm \mu \cos(\theta) \sqrt{2\left( \omega - I_{\mathrm{p}} \right)}.
\end{align}
The advantage of Eq.~\eqref{Phi1 model} is that it gives an analytical prediction of the first-order Stark phase directly in terms of the harmonic frequency, $\omega$, rather than in terms of electron trajectories through $\omega(t,t')$. We return to a discussion of the accuracy of this result below.

The second-order Stark phase cannot be expressed as simply in terms of the harmonic frequency. This is because Eq.~\eqref{Phi2} turns out to depend explicitly on the driving pulse, as well as giving qualitatively different result for short and long trajectories. The scaling of the second-order Stark phase with respect to laser parameters can be estimated by neglecting the pulse envelope, and integrating Eq.~\eqref{Phi2} from the peak of a half-cycle $t' = t_{\mathrm{peak}}$ up to $t = t_{\mathrm{peak}} + 2T/3$. The cut-off harmonics are then found to accumulate a second-order Stark phase proportional to $\rmF_0^2/\omega_0 = 4 \omega_0 U_{\mathrm{p}}$, where $\omega_0$ is the frequency of the driving laser, and $U_{\mathrm{p}}$ the ponderomotive potential. According to the usual cutoff rule, $\omega_{\mathrm{max}} = 3.17 U_{\mathrm{p}} + I_{\mathrm{p}}$, the importance of the second-order Stark phase can be reduced by increasing the wave length of the driving laser while keeping the harmonic cutoff fixed.

\begin{figure}
  \begin{center}
   \includegraphics[width=\columnwidth]{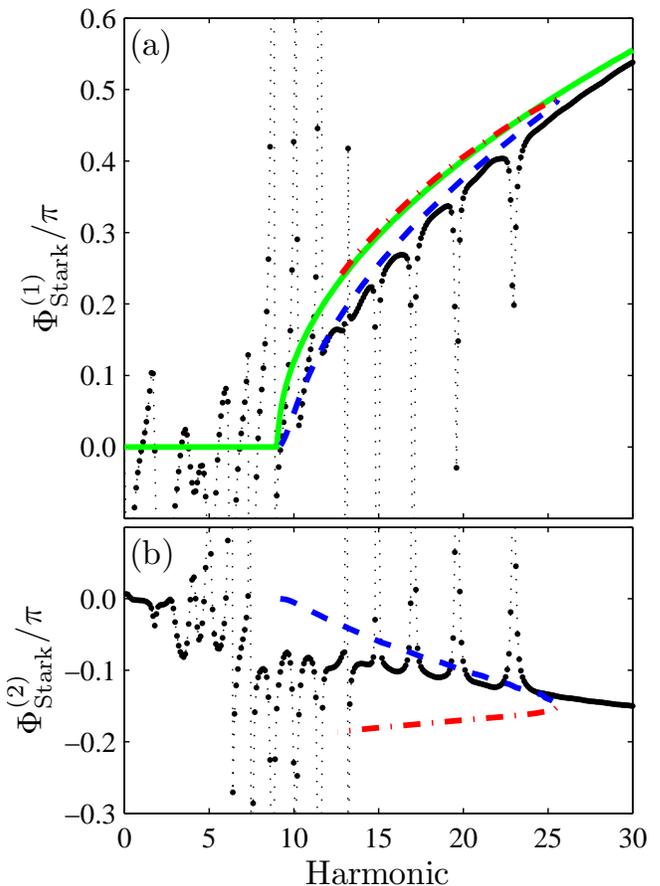}
 \end{center}
 \caption{(Color online) (a) Accumulated first-order Stark phase of CO subject to an ultrashort laser pulse (parameters are given in the text). The connected black dots indicate the phase obtained from a quantum calculation. The dashed (blue) curve is the phase predicted from short classical trajectories, and the dash-dotted (red) curve the phase from long trajectories. The smooth (green) curve is the analytical model of Eq.~\eqref{Phi1 model}, with the phase defined to be zero for harmonics below $I_{\mathrm{p}}$. (b) Accumulated second-order Stark phase. Dashed and dash-dotted curves indicate short and long classical trajectories.}
  \label{Fig3}
\end{figure}

We use the Lewenstein model~\cite{lewenstein:2117} to test the validity of the trajectory calculations. The first-order Stark phase is extracted by subtracting harmonic phases calculated with and without inclusion of the first-order Stark shift in the Lewenstein model~\cite{etches:155602, etches:023418}. Figure~\ref{Fig3}(a) shows that short trajectories account for the first-order Stark phase within the limits set by the classical model, which does not account for harmonics below the ionization threshold, nor beyond the harmonic cutoff. The phase spikes coincide with minima in the harmonic plateau. The minima are caused by the interference between short and long trajectories, and each is associated with a sharp variation in the harmonic phase. The exact position of each minimum changes when the Stark shift is included, which explains the phase spikes. 

The analytical model of Eq.~\eqref{Phi1 model} slightly overestimates the first-order Stark phase due to the use of the field-free ionization potential. The error is therefore expected to increase with increasing intensity. Comparing Eq.~\eqref{Phi1 model} to Lewenstein calculations at the 21st harmonic for three different intensities we observe an error of $8\%$ ($1.5 \times 10^{14}$~W$/$cm$^2$), $10\%$ ($2.0 \times 10^{14}$~W$/$cm$^2$), and $12\%$ ($2.5 \times 10^{14}$~W$/$cm$^2$). The interference features also move with intensity, but the square root behavior is the same at all three intensities. The close agreement between Eq.~\eqref{Phi1 model} and the long trajectory calculation is due to the extremely narrow pulse envelope, which causes the long trajectories to recombine at low field strengths as shown in Fig.~\ref{Fig2}.

The second-order Stark phase is plotted in Fig.~\ref{Fig3}(b). The Lewenstein result is obtained by including the first and second-order Stark shifts in the Lewenstein model, calculating the harmonic phases, and then subtracting the phase obtained when only the first-order Stark shift is included. The calculation agrees qualitatively with the short trajectory prediction. The trajectory calculation shows that the second-order Stark phase increases in magnitude when the electron spends longer time in the continuum. Selecting the short-trajectory contribution is thus an additional way of reducing the importance of second-order Stark phases.

We would like to stress the point that the simple behavior of the Stark phase is due to the fact that only one half-cycle, and mostly the short trajectories, contributes to the high-order harmonics. When several sets of trajectories contribute, the total phase is the result of a coherent sum of harmonic amplitudes, which can cause large modulations on top of the trend in Fig.~\ref{Fig3}. Experimentally, the dominating trajectory is selected through phase-matching by adjusting the position of the laser focus relative to the nonlinear medium~\cite{gaarde:132001}. 

In order to uncover the interplay between Stark phases and phase-matching, we now present results of the coupled solutions of the Maxwell wave equation (MWE) and the time-dependent Schr\"odinger equation (TDSE). We solve the MWE in the slowly evolving wave approximation as described in~\cite{gaarde:132001}, and at each plane in the propagation direction we solve the TDSE in the Lewenstein model to calculate the time-dependent dipole moment~\cite{etches:155602}. The nonlinear medium is a $5$~mm jet of CO molecules. The CO molecules are assumed to be perfectly oriented as in Fig.~\ref{Fig2}. The gas density is set to $5 \times 10^{14}$~cm$^{-3}$. We use the same driving pulse as above, except for adding a Gaussian focus with confocal parameter $b = 2.0$~cm. The focus is placed $0.70$~cm in front of the middle of the jet. The peak intensity is $3.0 \times 10^{14}$~W$/$cm$^2$, chosen so as to give a peak intensity of $2.0 \times 10^{14}$~W$/$cm$^2$ in the middle of the medium. Our results are insensitive to ionization due to the very low target density. 

\begin{figure}
  \begin{center}
   \includegraphics[width=\columnwidth]{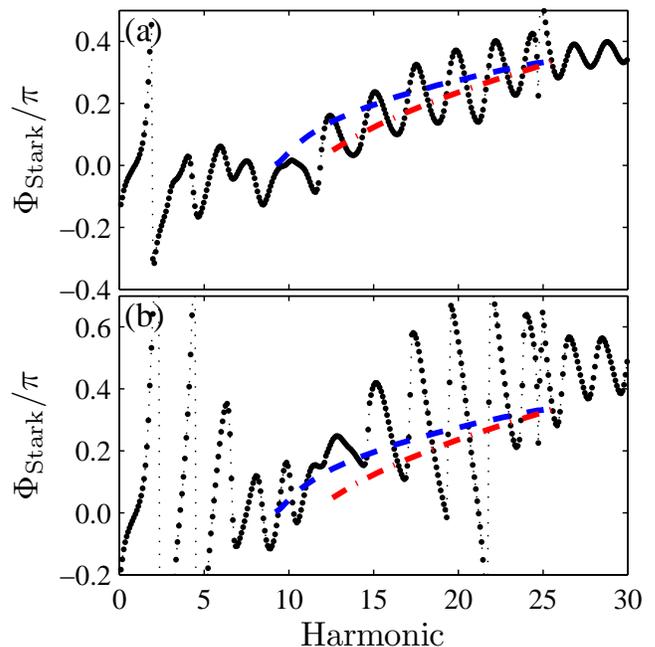}
 \end{center}
 \caption{(Color online) Stark phases from a gas jet of CO molecules (parameters are given in the text). (a)~The CO molecules are oriented as in Fig.~\ref{Fig2}. The connected black dots are phases obtained from a full simulation including macroscopic propagation. The dashed (blue) curve is the single-molecule phase predicted from short classical trajectories, and the dash-dotted (red) curve the phase from long trajectories. (b)~The CO molecules are aligned parallel to the laser polarization. The black dots include both orientations of the target molecules. Dashed and dash-dotted curves indicate short and long classical trajectories.}
  \label{Fig4}
\end{figure}

The Stark phase is calculated by propagating the MWE twice, with and without including the first- and second-order Stark effect in the single-molecule Lewenstein calculations. In each case we transform to the far field, apply a spatial filter that selects predominantly the central, short trajectory contribution to the harmonics, and transform back to the near field. In an experiment this would correspond to having an aperture or a refocusing mirror in the far field. Then we subtract the phases obtained with and without including Stark shifts. An average over the final spot on the detector screen is made by weighting the Stark phase $\Phi_{\mathrm{Stark}}(\omega, r)$ at a given radius with the strength of the harmonic $\left| \rmF_{\mathrm{HHG}}(\omega, r) \right|^2$:
\begin{align}
  \label{r average}
  \left< \Phi_{\mathrm{Stark}}(\omega) \right>_r = \frac{\int_0^{\infty} \Phi_{\mathrm{Stark}}(\omega, r) \left| \rmF_{\mathrm{HHG}}(\omega, r) \right|^2 r \, \rmd r}{\int_0^{\infty} \left| \rmF_{\mathrm{HHG}}(\omega, r) \right|^2 r \, \rmd r}.
\end{align}
The result, shown in Fig.~\ref{Fig4}(a), compares qualitatively with single-molecule predictions based on classical trajectories calculated for a peak intensity of $2.0 \times 10^{14}$~W$/$cm$^2$. The phase oscillations stem from the intensity dependence of minima caused by interference between short and long trajectories. At a fixed intensity the minima are very sharp, giving rise to sharp variations in the extracted Stark phase as in Fig.~\ref{Fig3}. In macroscopic calculations the intensity falls off along, and perpendicular to, the propagation axis of the driving laser, causing the minima to smear out. The Lewenstein model has been known to overestimate the importance of the long trajectories~\cite{gaarde:031406}, leading to an exaggeration of these interference oscillations in our calculations. 

We also present results for perfectly aligned CO molecules. Spectra are calculated with and without the Stark effect by solving the MWE separately for opposite orientations, and adding the resulting harmonics coherently at the end of the gas~\cite{madsen:035401}. This procedure is valid in the limit where ionization-induced reshaping of the driving pulse is neglible. The harmonics are then propagated to the far field, filtered, refocused, and the Stark phase extracted. Equation~\eqref{r average} is used to calculate the radially averaged Stark phase shown in Fig.~\ref{Fig4}(b). If the first-order Stark phase from opposite orientations had canceled out, then the total Stark phase would have been similar to that in Fig.~\ref{Fig3}(b). Instead, the total Stark phase is similar to that of fully oriented CO molecules in Fig.~\ref{Fig4}(a). The reason for this is that the Lewenstein model favors ionization when the electric field points from carbon to oxygen~\cite{etches:155602}. The ultrashort pulse only allows one half-cycle to contribute, thus increasing the relative contribution from the orientation shown in Fig.~\ref{Fig2}. The nonvanishing first-order Stark phase in Fig.~\ref{Fig4} underlines that polar molecules behave differently from nonpolar molecules, even if their head-to-tail symmetry is unbroken.

In conclusion, we have investigated the role of laser-induced bound-state phases in HHG. We show that first- and second-order Stark shifts may both contribute significantly to the phase of harmonics generated from polar molecules. Such Stark phases must be accounted for if HHG is to be used for orbital tomography. We find a simple analytical expression for the first-order Stark phase, showing it to scale as the square root of harmonic frequency. No simple expression is found for the second-order Stark phase, but it can be minimized by increasing the laser wave length while keeping the ponderomotive potential fixed. Propagation of the Maxwell wave equation confirms that Stark phases survive phase-matching in the target gas.

This work was supported by the Danish National Research Council (Grant No.~10-85430), the National Science Foundation under Grant No.~PHY-1019071, and an ERC-StG (Project NO.~277767 -TDMET). High-performance computational resources were provided by the Louisiana Optical Network Initiative, www.loni.org. We would like to thank Hans-Jakob W\"orner for advice regarding experimental parameters.


\end{document}